\documentclass[12pt]{article}
\pdfoutput=1
\usepackage{amssymb}
\usepackage{amsmath}
\usepackage{amsfonts}
\usepackage{graphicx}
\usepackage{mathrsfs}
\usepackage{comment}
\usepackage{cite}
\usepackage{color}
\usepackage{comment}

\newcommand{\mathhyphen}{\mathchar"712D}

\begin{document}

\renewcommand{\figurename}{Fig.}
\renewcommand{\tablename}{Table.}
\newcommand{\Slash}[1]{{\ooalign{\hfil#1\hfil\crcr\raise.167ex\hbox{/}}}}
\newcommand{\bra}[1]{ \langle {#1} | }
\newcommand{\ket}[1]{ | {#1} \rangle }
\newcommand{\bef}{\begin{figure}}  \newcommand{\eef}{\end{figure}}
\newcommand{\bec}{\begin{center}}  \newcommand{\eec}{\end{center}}
\newcommand{\non}{\nonumber}  \newcommand{\eqn}[1]{\begin{equation} {#1}\end{equation}}
\newcommand{\laq}[1]{\label{eq:#1}}  
\newcommand{\dd}[1]{{d \o d{#1}}}
\newcommand{\Eq}[1]{Eq.(\ref{eq:#1})}
\newcommand{\Eqs}[1]{Eqs.(\ref{eq:#1})}
\newcommand{\eq}[1]{(\ref{eq:#1})}
\newcommand{\ab}[1]{\left|{#1}\right|}
\newcommand{\vev}[1]{ \left\langle {#1} \right\rangle }
\newcommand{\bs}[1]{ {\boldsymbol {#1}} }
\newcommand{\lac}[1]{\label{chap:#1}}
\newcommand{\SU}[1]{{\rm SU{#1} } }
\newcommand{\SO}[1]{{\rm SO{#1}} }
\def\({\left(}
\def\){\right)}
\def\dt{{d \o dt}}
\def\diag{\mathop{\rm diag}\nolimits}
\def\Spin{\mathop{\rm Spin}}
\def\O{{O}}
\def\U{\mathop{\rm U}}
\def\Sp{\mathop{\rm Sp}}
\def\SL{\mathop{\rm SL}}
\def\tr{\mathop{\rm tr}}
\def\ebq{\end{equation} \begin{equation}}
\newcommand{\OR}{~{\rm or}~}
\newcommand{\AND}{~{\rm and}~}
\newcommand{\EV}{ {\rm \, eV} }
\newcommand{\KEV}{ {\rm \, keV} }
\newcommand{\MEV}{ {\rm \, MeV} }
\newcommand{\GEV}{ {\rm \, GeV} }
\newcommand{\TEV}{ {\rm \, TeV} }
\def\o{\over}
\def\a{\alpha}
\def\b{\beta}
\def\c{\varepsilon}
\def\d{\delta}
\def\e{\epsilon}
\def\f{\phi}
\def\g{\gamma}
\def\h{\theta}
\def\k{\kappa}
\def\l{\lambda}
\def\m{\mu}
\def\n{\nu}
\def\p{\psi}
\def\q{\partial}
\def\r{\rho}
\def\s{\sigma}
\def\t{\tau}
\def\u{\upsilon}
\def\w{\omega}
\def\x{\xi}
\def\y{\eta}
\def\z{\zeta}
\def\D{\Delta}
\def\G{\Gamma}
\def\H{\Theta}
\def\L{\Lambda}
\def\F{\Phi}
\def\P{\Psi}
\def\S{\Sigma}
\def\me{\mathrm e}
\def\ol{\overline}
\def\tl{\tilde}
\def\*{\dagger}

\begin{center}

\hfill November, 2020

\vspace{1.5cm}

{\LARGE\bf  Light Dark Matter from Inflaton Decay
}
\vspace{1.5cm}

{\Large Takeo Moroi and Wen Yin}

\vspace{12pt}
\vspace{1.5cm}
{\em 
  { Department of Physics, The University of Tokyo, Tokyo 113-0033, Japan}
}
\vspace{12pt}
\vspace{1.5cm}

\date{\today $\vphantom{\bigg|_{\bigg|}^|}$}

\abstract{
We propose a simple mechanism of light dark matter (DM) production
from the decay of the oscillating inflaton condensation.  If the
reheating temperature after inflation is higher than the inflaton
mass, which is of the same order of the momentum of the DM at the time
of the production, the DM momentum can be suppressed compared to the
temperature of the thermal plasma if the interaction of the DM is weak
enough.  Consequently, the DM can be cold enough to avoid the
observational constraints on the warm DM, like the Lyman-$\alpha$
bound even if the DM mass is small.  We study the bosonic and
fermionic DM production from the inflaton decay, taking into account
the effect of the stimulated emission and Pauli blocking,
respectively.  In both cases, the DM can be cold and abundant enough
to be a viable candidate of the DM.  We also apply our mechanism to
the production of isocurvature-problem-free axion DM and Dirac sea DM
of right-handed neutrino consistent the seesaw relation for the
active neutrino masses.

}

\end{center}

\clearpage

\renewcommand{\thefootnote}{\#\arabic{footnote}}
\setcounter{footnote}{0}

\setcounter{page}{1}

\setcounter{footnote}{0}

\section{Introduction}
\setcounter{equation}{0}

A long-standing puzzle of particle physics and cosmology is the origin
of dark matter (DM).  The DM is known to be (very) weakly coupled to
the standard model (SM) particles, stable, cold, and abundant in the
present Universe. However the mass, the interactions, and the
production mechanism are still not clear.

The DM mass may be small, in which case the stability (or longevity)
is easily explained due to the kinematics (or suppressed decay width).
From the point of view of DM direct detections, if the mass is small
enough, the recoil energy of a nucleon via the DM-nucleon scattering
is highly suppressed, which is consistent with the null result of the
direct detection experiments for WIMP.  Nevertheless a light DM can be
searched for from different approaches in the near future
(e.g.\ Ref.~\cite{Irastorza:2018dyq} for review of axion searches,
Refs.~\cite{Zhang:2018xdp, Akerib:2019fml, Aprile:2020tmw} for direct
detection with electron recoils).

A difficulty of the light DM is the production in the early Universe.
If it were produced thermally, like the WIMP, the number density
cannot be larger than that of the SM photons, and may be too small
unless the mass is larger than $\EV$.  Moreover, it may be too hot to
be consistent with structure formation.  For instance a thermal relic
of sterile neutrino has to be heavier than $2-5
\KEV$~\cite{Viel:2005qj, Irsic:2017ixq} from Lyman-$\a$ forest data.
If it is produced from freeze-in the bound is even
severer~\cite{Kamada:2019kpe}.

The production of light DM has been discussed widely. An
axion/axion-like particle (ALP) can be produced via the misalignment
mechanism~\cite{Preskill:1982cy,Abbott:1982af,Dine:1982ah}.  A light
hidden photon DM can be produced gravitationally \cite{Graham:2015rva, Ema:2019yrd}.  (Gravitational effect during inflation is also
important for the axion/ALP DM~\cite{Preskill:1982cy,Abbott:1982af,Dine:1982ah, Graham:2018jyp,Guth:2018hsa}).
The hidden photon DM can be produced from a parametric/tachyonic
resonance by coupling to the QCD
axion~\cite{Kitajima:2017peg,Agrawal:2018vin,Co:2018lka,Dror:2018pdh}
(see also Refs\,\cite{Agrawal:2017eqm}). In addition, the axion (or
ALP) production from the parametric resonance of another scalar field
was discussed~\cite{Mazumdar:2015pta, Co:2017mop}.  A light DM may be
produced by the decays of heavy particles at an early
epoch~\cite{Randall:2015xza}.

In this paper, we study a new possibility of light DM production.  The
light DM is produced from the decay of an inflaton which reheats the
Universe.\footnote
{The parent scalar field does not have to be the inflaton; if it once
  dominates the universe, any scalar field may play the role.  For the
  minimality of the scenario, in the this analysis, we assume that the
  parent scalar field is the inflaton.}
We show that, if the reheating temperature $T_R$ is higher than the
inflaton mass $m_\f$, the DM particles from the inflaton decays can
become cold enough due to the red shift.  If the inflaton coupling to
the SM particles and the DM is not too small, $T_R\gtrsim m_\f$ can be
realized due to the dissipation effect~\cite{Yokoyama:2005dv,
  Anisimov:2008dz, Drewes:2010pf, Mukaida:2012qn, Drewes:2013iaa,
  Mukaida:2012bz, Moroi:2014mqa}.  In particular, for the case that
the DM is bosonic, the DM production can be enhanced due to the
stimulated emission effect like LASER (light amplification by
stimulated emission of radiation); we call such a mechanism as DASER.
In addition, if the DM is fermionic, a Dirac sea DM can be realized.
Interestingly, the produced DM does not suffer from the isocurvature
problem.  Axion(-like) DM and Dirac sea DM of a right-handed neutrino,
which is responsible to the neutrino oscillations, are discussed.

Heavy DM production in association with the reheating was studied,
e.g.\ due to the decay (with a late time annihilation) of the inflaton~\cite{Moroi:1994rs, Kawasaki:1995cy, Moroi:1999zb, Jeong:2011sg, Ellis:2015jpg}, during thermalization at the last stage of the reheating
~\cite{Harigaya:2014waa, Garcia:2018wtq, Harigaya:2019tzu, Garcia:2020eof} and during
the preheating \cite{Chung:1998zb, Chung:1998ua}.  In contrast, we
focus on general ``light" DM production, for which we find that
$T_R/m_\f\gtrsim 1$ is important.  In Ref\,.\cite{Co:2017mop}, the QCD
axion production from parametric resonance of the Peccei-Quinn (PQ)
field was considered, however the PQ field condensate does not
dominate and reheat the Universe.

This paper is organized as follows.  In Section \ref{sec:formalisms},
we present the basic formalisms used in our analysis.  In Section
\ref{sec:LDM}, we discuss the light DM production from the inflaton
decay.  In Section \ref{sec:models}, we discuss several possibilities
of the light DM produced by our mechanism, i.e., axion DM and
right-handed neutrino DM.  In Section \ref{sec:M2}, we give an
example of the inflaton model which gives $T_R\gtrsim m_\phi$.
Section \ref{sec:conclusions} is devoted to conclusions and
discussion.

\section{Basic Formalisms}
\label{sec:formalisms}
\setcounter{equation}{0}

Throughout this paper we will use a Boltzmann equation, which governs
the evolution of the distribution function of the DM particle, to
describe the DM production from the inflaton decay.  In this section,
we introduce and solve the Boltzmann equation in expanding Universe to
discuss the DM abundance produced by the decay of a scalar field
$\phi$ (which is identified as the inflaton).  We also discuss some
relations between the Boltzmann equation and resonance parameter,
which is introduced for a microscopic picture.

Throughout this paper, we assume that the potential of $\phi$ is well
approximated by a parabolic one and that $\phi$ is spatially
homogeneous.  Then, for the timescale much shorter than the cosmic
expansion, the motion of $\phi$ is given by
\begin{align}
  \phi (t) = \bar{\phi} \cos m_\phi t,
  \label{phioscillation}
\end{align}
where $\bar{\phi}$ is the amplitude while $m_\phi$ is the mass of
$\phi$.  The number density of $\phi$ is given by
\begin{align}
  n_\phi = \frac{1}{2} m_\phi \bar{\phi}^2.
\end{align}

The Boltzmann equation of $\chi$ in the expanding Universe is given
as
\begin{align}
  \dot{f}_{\vec{k}}= H \vec{k} \frac{\partial f_{\vec{k}}}{\partial \vec{k}}+
  \dot{f}_{\vec{k}}^{\rm (coll)},
  \label{Boltzmanneq}
\end{align}
where $H$ is the expansion rate of the universe:
\begin{align}
  H = \frac{\dot{a}}{a},
\end{align}
with $a$ being the scale factor, $f_{\vec{k}}$ is a distribution
function of $\chi$ with three momentum $\vec{k}$, and
$\dot{f}_{\vec{k}}^{\rm (coll)} $ is the collision term (see, 
e.g.\ \cite{Kolb:1990vq}).  Hereafter, we concentrate on the case
that the DM production is via the two-body decay of the inflaton,
$\phi\rightarrow\chi\chi$.  In such a case,\footnote
{We consider the case that $n_\f$ is so large that we can omit ``1''
  in the Bose-enhancement factor for $\f$ in the inverse decay
  process.  It is also assumed that the change of $\bar{\phi}$ is so
  slow that $\bar{\phi}$ can be well approximated to be a constant in
  deriving collision term. } the collision term is given by
\begin{align}
  \dot{f}_{\vec{k}}^{\rm (coll)} =
  2 n_\phi \Gamma_{\phi\rightarrow\chi\chi}^{(0)}
  \left[ (1 \pm f_{\vec{k}}) (1 \pm f_{-\vec{k}}) - f_{\vec{k}} f_{-\vec{k}} \right]
  \delta( |\vec{k}| - p_\chi )
  \left( \frac{p_\chi^2}{2\pi^2} \right)^{-1},
  \label{CollisionTerm}
\end{align}
where, in the square bracket, $(1+f)$ and $(1-f)$ are for bosonic and
fermionic DM, respectively, and
$\Gamma_{\phi\rightarrow\chi\chi}^{(0)}$ is the decay rate.  The
superscript ``(0)'' indicates that $\Gamma_{\phi \rightarrow \chi
  \chi }^{(0)}$ is the perturbative decay rate at the vacuum.  In
addition, $p_\chi$ is the three-momentum of $\chi$ produced by the
decay:
\begin{align}
  p_\chi \equiv \frac{1}{2} m_\phi \sqrt{1-\frac{4m_\chi^2}{m_\phi^2}}.
\end{align}
The first and second terms in the square bracket in
Eq.\ \eqref{CollisionTerm} are understood as the contributions of the
decay and the inverse decay (taking into account the stimulated
emission or the Pauli blocking), respectively.\footnote
{Notice that, in Eq.\ \eqref{CollisionTerm}, the factor of $2$
  in the right-hand side reflects the fact that two $\chi$'s are
  produced by the decay of single $\phi$.  One can see that, for
  $f_{\vec{k}}\ll 1$,
  \begin{align*}
    \dot{n}_\chi^{\rm (coll)} \equiv
    g \int \frac{d^3 k}{(2\pi)^3} \dot{f}_{\vec{k}}^{\rm (coll)}
    \simeq
    2 g n_\phi \Gamma_{\phi\rightarrow\chi\chi}^{(0)}.
  \end{align*}
}
Using the distribution function, the number density of $\chi$ is given
by
\begin{align}
\label{number1}
  n_\chi (t) = g \int \frac{d^3k}{(2\pi)^3} f_{\vec{k}} (t),
\end{align}
with $g$ being the multiplicity.  If $\chi$ is a real scalar and a
singlet Weyl fermion, $g=1$ and $2$, respectively. When $g> 1$,
$\G_{\f \to \chi \chi}^{(0)}$ is defined as the decay rate to a single
$\chi$ pair; it is not the decay rate to $g$ pairs of $\chi$.

Now, we estimate the DM abundance from the decay of $\phi$, taking
into account the effects of the cosmic expansion.  In the following,
the distribution of $\chi$ is assumed to be isotropic so that the
distribution function does not depend on the direction of $\vec{k}$
but depends only on $k\equiv |\vec{k}|$.  Hereafter, the distribution
function is denoted as $f_k$.  In solving the Boltzmann equation in
the expanding universe, it is convenient to introduce the comoving
momentum:
\begin{align}
  \hat{k} (k) \equiv \frac{a(t)}{a_0} k,
\end{align}
where $a_0$ is the scale factor at some reference time.  Let us
define:
\begin{align}
  \hat{f}_{\hat{k}} (t) \equiv f_{k} (t).
\end{align}
Then, $\hat{f}_{\hat{k}}$ satisfies the following differential
equation:
\begin{align}
  \dot{\hat{f}}_{\hat{k}} =
  \frac{2 n_\phi \Gamma_{\phi \to \chi \chi}^{(0)}}{H p_\chi}
  (1\pm 2\hat{f}_{\hat{k}})
  \delta( t - t_{\hat{k}} )  
  \left( \frac{p_\chi^2}{2\pi^2} \right)^{-1}.
  \label{dot(hat(f))}
\end{align}
Here, $t_{\hat{k}}$ is the cosmic time when $k$ becomes equal to
$p_\chi$ for a given $\hat{k}$, i.e., it is given by
\begin{align}
  a (t_{\hat{k}}) = a_0 \frac{p_\chi}{\hat{k}}.
\end{align}
Solving Eq.\ \eqref{dot(hat(f))} with $f_k (t\rightarrow 0)=0$, we
obtain\footnote
{When $\bar{f}\gg 1$, in fact, a quantum effect may become important,
  which reduces the exponent by a factor of $1/2$.  For more
  discussion about the particle production in the framework of the
  quantum field theory, as well as the relation between the parametric
  resonance and the Bose-enhancement factor in the Boltzmann equation,
  see \cite{Moroi:2020bkq}.  Although the quantum effect is important
  for a precise calculation of the DM abundance, it does not affect
  our discussion in the following sections.}
\begin{align}
  f_k (t\rightarrow\infty) =
  \pm \frac{1}{2}
  \left( e^{\pm 2\bar{f}(t_k)} - 1 \right)
  \theta (p_\chi - k),
  \label{f_k(exp)}
\end{align}
with
\begin{align}
\laq{expo}
  \bar{f}(t_k) \equiv
  \left.  \frac{4\pi^2 \Gamma_{\phi \to \chi \chi}^{(0)} n_\phi}{H p_\chi^3} \right|_{t=t_{\hat{k}}}.
\end{align}

For the case of bosonic DM, the production of the DM is exponentially
enhanced when $\bar{f}\gtrsim 1$ due to the stimulated emission. This
is like the light amplification by stimulated emission of radiation,
LASER; we call the DM amplification by stimulated emotion of radiation
(i.e., relativistic DM) as DASER.

Before closing this section, we comment on the Boltzmann equation from
the point of view of parametric resonance.  To this end, we assume the
interacting Lagrangian
\begin{align}
{\cal L}_{\rm int}=- A \phi \chi^2,
  \label{Hphichichi}
\end{align}
for the interaction of $\phi$ with a real scalar field $\chi$.  Then
the decay rate is obtained as
\begin{align}
  \Gamma_{\phi \rightarrow \chi \chi }^{(0)} =
  \frac{A^2}{8\pi m_\phi} \frac{2p_\chi}{m_\phi}.
  \label{Gamma(phi->chichi)}  
\end{align}
Further, for simplicity, we neglect the mass of $\chi$, and thus
$m_\phi \simeq 2p_\chi$.

The production of $\chi$ in this system has been also studied in the
context of parametric resonance \cite{Kofman:1994rk, Kofman:1997yn, Dufaux:2006ee, Amin:2018kkg}.
In the particle production through the parametric resonance, the
particles in resonance bands are effectively produced.  The widths of
the bands are typically of $O(qm_\phi)$, where
\begin{align}
  q \equiv \frac{4A\bar{\phi}}{m_\phi^2}.
  \label{q-def}
\end{align}
We study the case that the band widths are narrow, i.e.,
\begin{align}
  q \ll 1.
\end{align}
Then, the DM particles are produced almost ``on-resonance,'' i.e., the
momentum of the DM produced from the inflaton oscillation is
$|\vec{k}|\simeq p_\chi$. Otherwise, the DM acquires an oscillating
mass with amplitude larger than $m_\f$ and the aforementioned decay
process is sometimes forbidden.

A given comoving momentum stays in the resonance band for the
timescale of $O(q/H)$.  Thus, for the validity of our argument, the
change of the inflaton amplitude during such a timescale should be
negligible, which requires
\begin{align}
  \frac{\dot{\bar{\f}}}{\bar {\f}} \ll q^{-1} {H}.
  \label{phibardot}
\end{align}

We also note here that the exponential growth shown in
Eq.\ \eqref{f_k(exp)} is consistent with the behavior observed in the
analysis of parametric resonance in the narrow resonance regime,
$q\ll1$~\cite{Kofman:1997yn}.  In fact, there exist processes which
are higher order in ${\cal L}_{\rm int}$.  So far, we have considered
the effects of $O(A^2)$; in the perturbative picture, they take
account of the decay and inverse decay processes
$\phi\leftrightarrow\chi\chi$.  Effects which are higher order in $A$
are due to scattering processes, in which a number of $\phi$'s
annihilate into $\chi$'s.  Such effects are suppressed when the
amplitude of $\phi$ is small enough; the higher order terms are
suppressed by powers of $q$ and are negligible
when $q\ll 1$ \cite{Matsumoto:2007rd}.

\section{Light DM from Inflaton Decay}
\label{sec:LDM}
\setcounter{equation}{0}

\subsection{Cold DM from inflaton decay}

Now we consider the DM production from the inflaton decay,
$\phi\rightarrow\chi\chi$.  The DM is assumed to have negligible
interaction rate with itself and with Standard Model (SM) particles.\footnote
{This condition may be favored because the DM should be cold and
  abundant throughout the thermal history, e.g.\ they should not be
  heated by the ambient SM plasma or by
  itself~\cite{Carlson:1992fn}. }
The inflaton, on the other hand, not only couple to the DM but also
couples to some SM particles. The latter coupling leads to the
reheating of the Universe.

The energy density of the universe is once dominated by that of the
inflaton oscillation after inflation/preheating.  Here,
we assume that the inflaton potential is well approximated by the
quadratic one in such a period (c.f. \cite{Khlebnikov:1996mc, Micha:2002ey, Micha:2004bv, Daido:2017wwb, Daido:2017tbr, Choi:2019osi, Takahashi:2020uio}).  In such a case, the inflaton energy
density $\rho_\phi$ scales as $a^{-3}$ during the period of inflaton
oscillation.  We denote the cosmic time at the beginning of the
inflaton oscillation as $t_{ i}$, and that at the completion of the
reheating as $t=t_R$.  (Thus, at $t=t_R$, energy density of the
inflaton is converted to that of radiation and the radiation-dominated
universe starts.)  In the following, for simplicity, we adopt the
sudden-decay approximation, i.e., the inflaton is approximated to
decay at $t=t_R$ for the estimation of the reheating temperature.

During the oscillating period, the inflaton mass should be larger than
the expansion rate of the universe $H$.  Then, based on the inequality
$m_\phi \gg H(t_R)\sim \sqrt{g_\star \pi^2/90}T_R^2/M_{\rm pl}$ (with
$g_\star$ being the effective number of relativistic degrees of
freedom), we obtain a constraint on the reheating temperature:
\begin{align}
  T_R\lesssim T_R^{\rm (max)}
  \equiv
\({\frac{g_\star \pi^2}{90}}\)^{-1/4} \frac{M_{\rm pl}}{r}
  \simeq 10^{12}\GEV \times \left( \frac{r}{10^6}\right)^{-1},
  \label{osccon}
\end{align}
where
\begin{align}
  r \equiv \frac{T_R}{m_\phi}.
\end{align}
Notice that $T_{R}^{\rm (max)}$ is defined as the maximal possible
reheating temperature for a fixed value of $r$.

We introduce the parameter $n_\phi^{(0)}(t_R)$ such that the energy
density of the inflaton for $t_i\ll t\ll t_R$ is well approximated
as
\begin{align}
  \rho_\phi (t) \simeq m_\phi n_\phi^{(0)}(t_R)
  \left( \frac{a(t_R)}{a(t)} \right)^3.
\end{align}
Then, we define the effective number of DM produced by one inflaton
as
\begin{align}
  B \equiv \frac{n_\chi (t_R)}{n^{(0)}_\phi (t_R)}.
\end{align}
Here, $n_{\chi }(t_R)=a(t_R)^{-3} \int_{t_{\rm inf}}^{t_R}{dt
  n_{\f}(t) a(t)^3\Gamma_{\f\to \chi}(t)}$, where $\Gamma_{\f\to
  \chi}(t)$ is the number density transferring rate, and $t_{\rm inf}$
is the time at the end of inflation.  This formula includes the
production during the preheating epoch.  Using $B$, the number density
of DM for $t\gtrsim t_R$ is given by
\begin{align}
  n_\chi (t) = B n^{(0)}_\phi (t_R)
  \left( \frac{a(t_R)}{a(t)} \right)^3.  
\end{align}
Notice that, as we will discuss in the following, the DM density may
be dominated by those produced at the beginning of the oscillating
period.  In such a case, $B$ can be much larger than the branching
fraction $Br(\phi\rightarrow\chi\chi)$.  Using the fact that the
entropy density $s$ scales as $a^{-3}$ after the reheating, we can
estimate the density parameter of DM as
\begin{align}
  \Omega_\chi \sim
  \frac{m_\chi n_\chi (t_R)}{s(t_R)}
  \left( \frac{\rho_{\rm crit}}{s_0} \right)^{-1}
  \sim \frac{3}{4}  m_\chi r B \frac{s_0}{\rho_{\rm crit}},
  \label{abundance}
\end{align}
where $s_0$ is the present entropy density and $\rho_{\rm crit}$ is
the critical density.  Numerically, we obtain
\begin{align}
  m_\chi \sim 0.26\EV \times \frac{1}{ r B}
  \left( \frac{\Omega_\chi h^2}{0.12} \right),
  \label{abcon}
\end{align}
with $h$ being the Hubble constant in units of $100\ {\rm km/sec/Mpc}$.

Next, let us consider the coldness of DM.  In the following, for
simplicity, we assume that the reheating is instantaneous, i.e.,
$t_R\sim t_i$.  (For the case of $t_R\gg t_i$, $\chi$ can be also
produced at earlier epoch and is colder due to an extra red-shift.)
In addition, we consider the case that $m_\chi\ll m_\phi$.

At the time of the inflaton decay (i.e., $T\sim T_R$, with $T$ being
the cosmic temperature), the produced DMs are relativistic and their
momenta are typically
\begin{align}
  P_\chi (T_R) \sim \frac{1}{2} m_\phi,
\end{align}
where we have assumed that the DM production is dominated at $t\sim
t_R$.  Notice that, if the DMs are mostly produced at $t\ll t_R$,
$P_\chi (T_R)$ is more suppressed and DMs become colder.  With the
expansion of the universe, the momentum is red-shifted as
$P_{\chi}(T)=\frac{g_{\star,s}(T)^{1/3} T}{g_{\star,s}(T_R)^{1/3}T_R}
P_{\chi} (T_R)$ (with $g_{\star,s}$ being the effective number of
relativistic degrees of for the entropy density), and hence the
velocity of the DM is estimated as
\begin{align}
  v (T) \sim \mbox{min}\,
  \left( \frac{g_{\star,s}(T)^{1/3} T}{2g_{\star,s}(T_R)^{1/3}m_\chi} r^{-1}, 1 \right).
  \label{vel}
\end{align}
If the DM mass is too small, the DM velocity becomes too large to be
consistent with the Lyman-$\alpha$ bound.  Here, we adopt the analysis
on a warm DM of sterile neutrino~\cite{Viel:2005qj, Irsic:2017ixq} to
estimate the lower bound on the DM mass.  Regarding the constraint on
the root mean square velocity of the warm sterile neutrino DM
$\sqrt{\vev{v_{\rm warm}(T)^2}}$ as that on $v(T)$ of our model, we
estimate the lower bound of the DM mass. (See a similar approach
to the FIMP dark matter \cite{Kamada:2019kpe}.)  Then, we obtain
\begin{align}
  m_\chi\gtrsim 0.9\KEV r^{-1}\(\frac{107.75}{g_{\star,s}(T_R)}\)^{1/3},
  \label{velcon}
\end{align}
where, to be conservative, we adopt the lower bound on the sterile
neutrino mass of $\sim 2\, {\rm keV}$ \cite{Viel:2005qj}. In the
analysis in Sec.\,\ref{sec:narrow}, it is not always the case that the
DM production is dominated when $t\sim t_R$, and hence we
will calculate the root mean square velocity using the actual DM
spectrum to derive the bound.

\begin{figure}
  \begin{center}  
    \includegraphics[width=145mm]{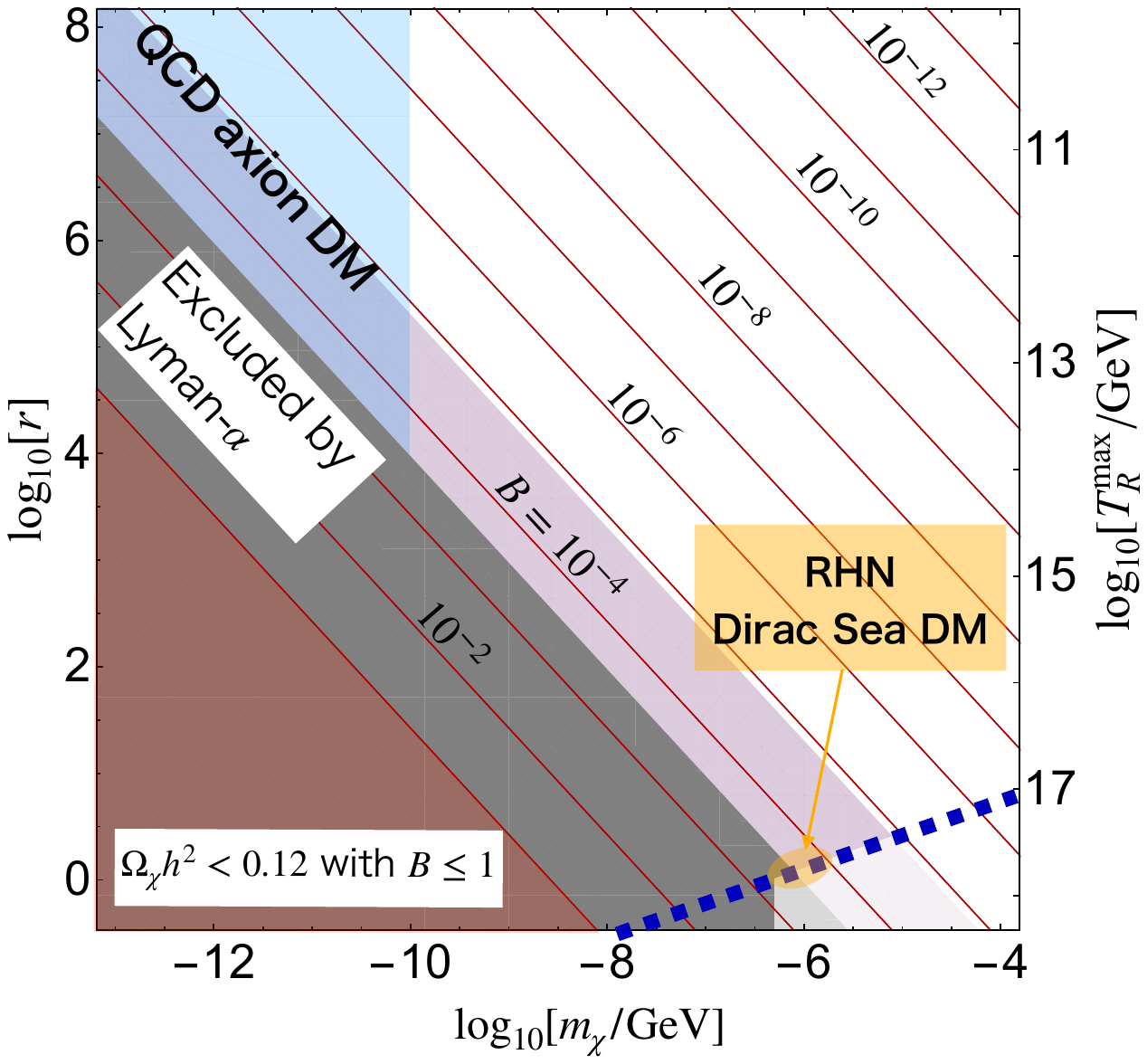}
  \end{center}
  \caption{The contour of constant $B$ to realize the present DM
    density on $m_\chi$ vs.\ $r$ plane (see Eq.\ \eqref{abcon}),
    taking $\Omega_\chi h^2=0.12$.  The maximal possible reheating
    temperature as a function of $r$ is shown on the right axis.
    In the gray region, the DM may be too hot to be consistent with
    Lyman-$\alpha$ forest data by assuming that the DM momenta at
    $t=t_R$ is typically $m_\phi/2$.  The purple region may be tested
    in the future.  For the case of the Fermionic DM, only the region
    below the blue-dotted line is relevant because of the
    Pauli-blocking (see Section \ref{sec:narrow}).  The typical
    parameter range of QCD axion DM and right-handed neutrino DM are
    shown in light blue and orange (see Section \ref{sec:model}).}
  \label{fig:DM} 
\end{figure}

When the DM is fermionic, there is a so-called Tremaine-Gunn (TG)
bound~\cite{Tremaine:1979we, Boyarsky:2008ju}:
\begin{align}
  m_\chi\gtrsim 0.5\KEV \(\frac{2}{g}\)^{1/4},
  \label{TG}
\end{align}
which is derived from the upper-bound on the phase space density in
dwarf spheroidal galaxies.  For $\chi$ forming a Dirac sea,
this bound is comparable to the recast Lyman-$\a$
bound~\cite{Randall:2016bqw}.

From Eqs.\ \eqref{abcon} and \eqref{velcon}, we get the key
observation that, even if the DM has a small mass, the coldness of the
DM can be realized with avoiding the Lyman-$\a$ bound taking $r\gtrsim
1$.  We emphasize that $r\gtrsim 1$ is easily realized if the
inflaton-SM coupling is large enough. In particular, $r\gg1$ can be
obtained when a dissipation effect, which is a scattering process
between the inflaton condensate and the SM plasma, is important to
complete the reheating~\cite{Yokoyama:2005dv, Anisimov:2008dz,
  Drewes:2010pf, Mukaida:2012qn, Drewes:2013iaa, Mukaida:2012bz,
  Moroi:2014mqa}. A general prediction in this dissipation regime is
that the inflaton is thermalized at the end of the reheating epoch.
The dissipation effect is rather generic and our conclusion below does
not depend on the detail of the model.  However, we provide a concrete
model in Section \ref{sec:model}.

The discussion so far is summarized in Fig.\,\ref{fig:DM} on $m_\chi$
vs.\ $r$ plane.  In Fig.\,\ref{fig:DM}, the contours of constant $B$
which satisfy the DM constraint given in Eq.\ \eqref{abcon} are
shown.  On the right axis, the aforementioned model-independent
maximal reheating temperature, $T_R^{\rm (max)}$, is shown.  The
recast Lyman-$\a$ bound \eqref{velcon} is shown as gray-shaded region,
while the current DM abundance cannot be realized in the red-shaded
region if $B\leq 1$.  One can see that the recast Lyman-$\a$ bound
fully excludes the parameter region with $B\geq 1$ and hence we do not
consider such a value of $B$.  The purple region may be tested by the
future 21cm line observation~\cite{Sitwell:2013fpa}; the region
corresponds to the warm sterile neutrino DM with $m_{N}\lesssim
20\ {\rm keV}$.  Thus, if $r$ is large enough, the DM produced by the
inflaton decay can be cold and abundant enough to be consistent with
observations.  We note here that all the region in the figure is
applicable to a bosonic DM; for bosonic DM, the effect of the Bose
enhancement becomes significant in the region above the blue dotted
line.  For the case of a fermionic DM, on the contrary, only the
region below the blue dotted line is relevant, as we will discuss in
the following.

\subsection{DM production in a narrow resonance regime}
\label{sec:narrow}

Now let us estimate the parameter $B$.
Given an inflation model and inflaton coupling to $\chi$, we can
estimate $B$.

In general, $B$ can be expressed as
\begin{align}
  B=B^{\rm (preh)}+B^{\rm (reh)}, 
\end{align}
where $B^{\rm (preh)}$ and $B^{\rm (reh)}$ are contributions during
the preheating (i.e., $t\lesssim t_i$) and during the inflaton
oscillation (i.e., $t_i \lesssim t\lesssim t_R$), respectively.
$B^{\rm (preh)}$ contributes to the IR mode of the DM distribution
function $f_k$.  The preheating epoch is highly model dependent.  For
instance, in the inflation models such as small-field inflation
  and hybrid inflation, tachyonic preheating generally takes place and
  the inflaton zero mode settles into the quadratic regime within
  $O(1)$
  oscillation~\cite{Felder:2000hj,Felder:2001kt,Brax:2010ai}.\footnote
{In this case, there are also produced spatial inhomogeneous modes,
  which have comparable energy to the inflaton condensate and are
  non-relativistic~\cite{Felder:2000hj,Felder:2001kt,Brax:2010ai}.
  The decays of the inhomogeneous modes may slightly modify the
  spectra, $f_k$, but we expect our estimation does not change much.
  Also, the field value of $\f$ just after inflation is far away from
  the minimum. If we use the interaction given in
  Eq.\ \eqref{Hphichichi}, the DM may either receives a heavy positive
  or negative mass squares depending on the sign of $A$. The latter
  case may generate the DM via another tachyonic preheating. This
  should correspond to a hybrid inflation.
  \label{ft:1}}
Hereafter we do not consider $B^{\rm (preh)}$, and concentrate on the
effects after the inflaton oscillation becomes effective.  Thus, we
mostly consider the case of $B\sim B^{(\text{reh})}$.

Now we estimate the DM abundance produced during the inflaton
oscillation.  The production in this epoch is somewhat
model-independent if the system is in narrow resonance regime, i.e.,
$q\ll1$.  Using the results given in Section \ref{sec:formalisms}, the
number density of the DM at time of the reheating is estimated as
\begin{align}
  n^{\rm (reh)}_{\chi} (t_R) \sim \pm \frac{1}{2}g \int^{p_{\chi}}_{k_{\rm IR}}
  \frac{d^3k}{(2\pi)^3}\(e^{\pm 2\bar{f}(t_k)}-1\),
\label{number}
\end{align}
where $k_{\rm IR}$ represents the momentum of DM produced at $t\sim
t_i$, i.e., $k_{\rm IR}\sim \frac{a(t_i)}{a({t_R})}p_\chi$.  If
$\bar{f}\ll1$ during the whole reheating epoch, we obtain $B^{\rm
  (reh)}\sim 2 t_R \Gamma_{\phi \to \chi \chi}^{(0)}$.  This result
does not depend whether the DM is a fermion or a boson.

\begin{figure}[!t]
  \begin{center}  
    \includegraphics[width=145mm]{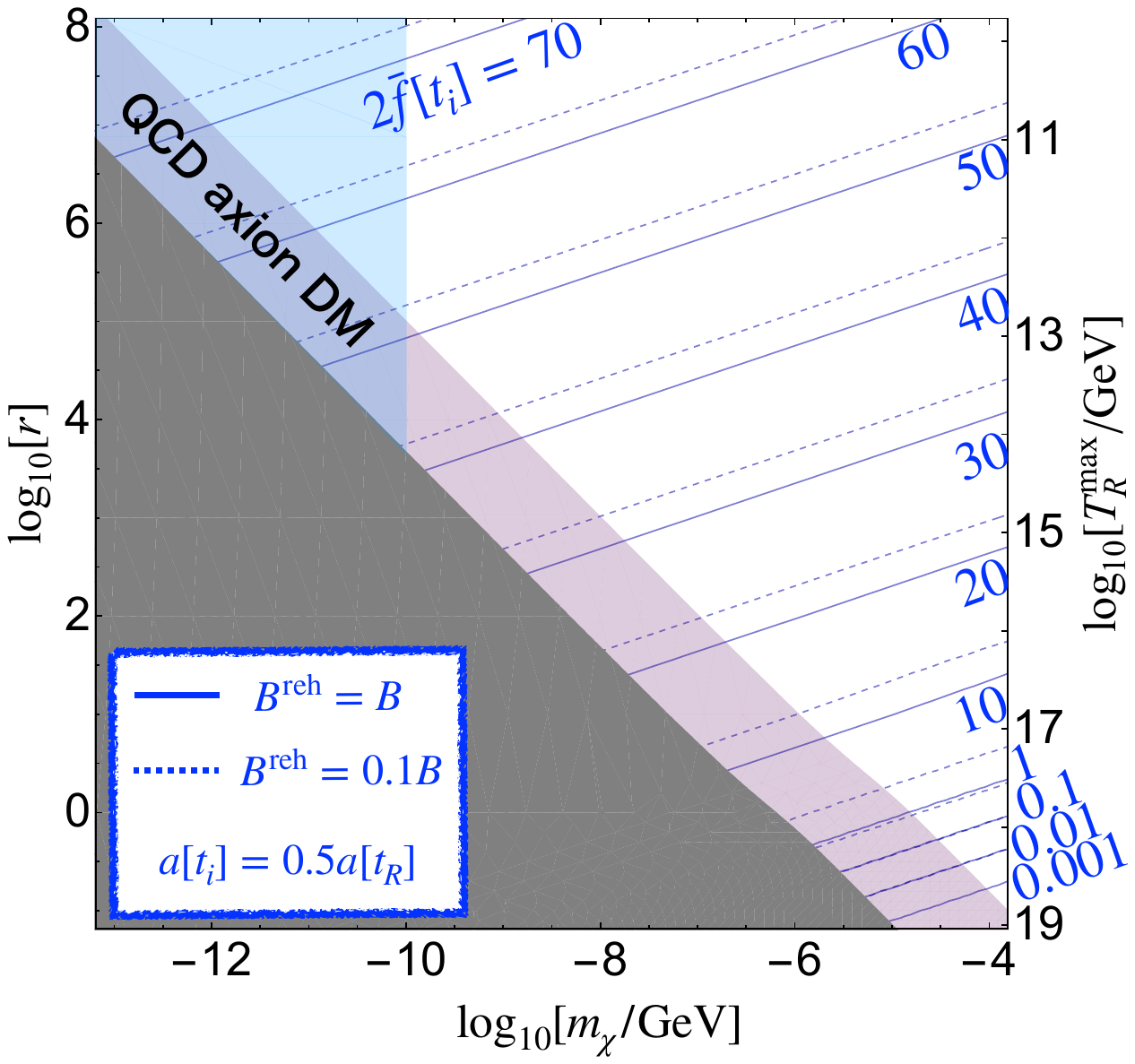}
  \end{center}
  \caption{ The contour of $2\bar{f}(t_i) $ of a bosonic DM in
    $m_\chi$-$r$ plane by requiring $\Omega_{\chi} h^2=0.12$ with
    $B^{\rm (reh)}=B$ (blue solid) and $B^{\rm (reh)}=0.1B$ (blue
    dotted).  We assume $a(t_i)=0.5a(t_R)$.  See Fig.\,\ref{fig:DM}
    for the meanings of the color-shaded regions. }
  \label{fig:fbar}
\end{figure}

The case with $\bar{f}\gtrsim 1$ is interesting.  In this case, the
exponential nature in Eq.\ \eqref{number} becomes important and $B$
depends on the statistics of $\chi$.  When $\chi$ is bosonic, the
integral is dominated at $k\sim k_{\rm IR}$; in other words, the
production of the DM becomes most efficient at the beginning of the
reheating at which the effect of the Bose enhancement becomes the
largest because $\bar{f}\propto n_\f/H \propto a^{-3/2}$.  Due to the
exponential enhancement, $B^{\rm (reh)}\gtrsim 2t_R \Gamma_{\phi \to
  \chi \chi}^{(0)}$.  Thus, the DM production during the reheating era
is enhanced via the DASER mechanism.  Since $e^{2\bar{f}}$ changes
rapidly as a function of time, $\bar{f}$ does not have to be much
larger than unity to obtain the value of $B$ given in
Fig.\ref{fig:DM}.  In Fig.\ref{fig:fbar}, we show the contours of
constant $2\bar{f}(t_i)$ to satisfy Eq.\ \eqref{abcon}, assuming
$a(t_i)=0.5a(t_R)$.  On the blue solid and dashed contours, $B^{\rm
  (reh)}=B$ and $B^{\rm (reh)}=0.1B$ are assumed, respectively.  In
the latter case, the contribution from the preheating epoch may be
dominant.  Notice that, for the case of $a(t_i)=0.5a(t_R)$, the red
shift during the inflaton oscillation period is short and
$n_\chi(t_R)\sim O(gp_\chi^3 e^{2\bar{f}(t_R)})$.  Then, we can find
that the value of $e^{2\bar{f}(t_R)}$ giving rise to the relevant DM
density scales as $\propto (r^3/m_\chi)$.  Numerically, for the case
of $\bar{f}(t_i)\gtrsim 1$, we find that Fig.\ \ref{fig:fbar} gives:
\begin{align}
  e^{2 \bar{f}(t_i)} \sim 3\times 10^{23}\ (3\times 10^{22})
  \times
  \left( \frac{r}{10^6} \right)^{3}
  \left( \frac{m_\chi}{10^{-8}\ {\rm GeV}} \right)^{-1},
 \label{fitfbar}
\end{align}
to make $\Omega_\chi$ to be consistent with the DM density parameter
with $B^{\rm (reh)}=B$ ($B^{\rm (reh)}=0.1B$).  This fit is relevant
for the parameter region with $r\gtrsim 1$.  We find that our
prediction of $\bar{f}$ does not change much even if the
contribution from the preheating epoch is sizable.\footnote
{If the inflaton condensate is significantly fragmented during the
  preheating (see footnote\,\ref{ft:1}), we may need $\bar{f}(t_i)$
  larger than the value given in Eq.\ \eqref{fitfbar} because
  the 
  exponential enhancement may not as strong as inflaton
  decay.  However, a certain level of the
  exponential growth is needed since the DM number within the sphere of radius
  $m_\f/2$ is over occupied to explain the DM abundance.  Thus we
  expect again that $2\bar{f}$ does not have to be much larger
  to realize the relevant DM density.}
We also use the distribution function given in Eq.\ \eqref{number} to
numerically estimate the averaged velocity, $\sqrt{\vev{v^2}}$, with
$B=B^{\rm (reh)}$ to recast the Lyman-$\a$ constraint and the future
reach of 21cm line observation, which are shown in the figure by the
gray and pink regions, respectively.

For the case that the DM is fermionic, on the other hand, the
Pauli-blocking effect should be taken into account if
$\bar{f}\gtrsim1$. Thus, the number density of the fermionic DM has an
upper-bound:
\begin{align}
  n_{\rm fermion\chi}^{\rm (max)} (t_R)
  = g \int_{k\leq p_\chi} \frac{d^3 k}{(2\pi)^3}
  =\frac{g m_\phi^3}{96\pi^2}.
\end{align}
This leads to 
\begin{align}
  B\leq \frac{m_\f n_{\rm fermion\chi}^{\rm (max)}(t_R)}{\rho_\f(t_R)}
  \sim 6\times 10^{-4}  r^{-4} g \(\frac{11}{g_\star}\)
  ~~:~~\mbox{fermionic DM}.
  \label{Fermimass}
\end{align} 
In Fig.\,\ref{fig:DM}, the above constraint is shown by the blue
dotted line.  In addition, combining Eq.\ \eqref{Fermimass} with
Eq.\ \eqref{abcon}, we obtain the DM mass to realize the Dirac sea DM
as
\begin{align}
  m_\chi\sim 0.4 \KEV  r^3 \(\frac{2}{ g}\) \(\frac{g_{\star, s}}{11}\)
  ~~:~~ \mbox{Dirac sea DM}.
\end{align}

So far, we have not considered back-reaction from plasma effect in the
$\chi$ production because the interaction of $\chi$ with the SM
particles or $\chi$ are assumed to be negligible.  Also, the plasma
effect induced by interaction between $\chi$ and thermalized $\phi$ is
suppressed due to the small coupling in the narrow resonance regime,
e.g.\ with the Lagrangian \eqref{Hphichichi}, the thermal correction
in the dispersion relation of $\chi$ should be $\d m_\chi^2\sim
\O(A^2)\ll q^2 m_\f^2.$

\section{DM Models}
\label{sec:models}
\setcounter{equation}{0}

\label{sec:model}

In this section, we consider two DM models in which our mechanism
plays an important role, i.e., QCD axion and right-handed neutrino
(RHN).

\subsection{DASER production of isocurvature free axion DM}
\label{sec:axion}

We first consider axion (-like) particle $a$.  (Thus, in this
subsection, the DM particle is denoted as $a$.)  The well-known
production mechanism of the axion DM is the misalignment
mechanism~\cite{Preskill:1982cy,Abbott:1982af,Dine:1982ah}.  That is,
when the (effective) axion mass becomes comparable to the expansion
rate of the universe, axion starts to oscillate around the potential
minimum. The coherent oscillation of axion becomes the cold DM.
However, if the PQ breaking scale is small, the initial
amplitude is suppressed and the resulting abundance, $\Omega^{\rm
  (mis)}_a,$ is suppressed.  If the PQ breaking happens before the
inflation, the quantum fluctuation of the axion may also contribute to
the power spectrum of isocurvature perturbation.  Our mechanism is
complementary to the misalignment mechanism and it works even when
$\Omega^{\rm (mis)}_a$ is smaller than the density parameter of
the DM.

In the following, we will show that the DASER production during reheating
may provide an alternative mechanism to produce cold axion DM.  For
concreteness, we adopt the following higher-dimensional operator for
the inflaton coupling to axion:
\begin{align}
  {\cal L}_{\rm int} = -\frac{\f}{\L_a} \partial_\mu a \partial^\mu a,
  \label{Hint(axion)}
\end{align}
where $\Lambda_a$ is a constant.  The higher dimensional operator
given above can originate from the coupling $\frac{2\f}{\L_a}
\partial_\mu\Phi_{\rm PQ}^* \partial^\mu\Phi_{\rm PQ}$, where
$\Phi_{\rm PQ}$ is the PQ field responsible for the breaking of the
$\U(1)_{\rm PQ}$ symmetry.  Alternatively, the $\f\to aa$ decays can be possible through the mixing between $\f$ and the PQ Higgs boson via the renormalizable coupling $ \f |\F_{\rm PQ}|^2$.
In both cases, the
$\f\text{-}{a}\text{-}{a}$ coupling is obtained by integrating out the
heavy PQ Higgs boson which is assumed to be heaver than $\f$.  Even
though the interaction given in Eq.\ \eqref{Hint(axion)} has a
different form from that in Eq.\ \eqref{Hphichichi}, the discussion on
resonance parameters in previous sections are applicable with
replacing $A\rightarrow\frac{m_\phi^2}{2\Lambda_a}$.  Then, the decay
rate of $\phi$ into the axion pair is given by
\begin{align}
  \Gamma^{(0)}_{\phi\rightarrow aa} =
  \frac{1}{32\pi} \frac{m_\phi^3}{\Lambda_a^2}.
\end{align}

Although our mechanism works for a large class of axion models, we 
focus on a QCD axion DM.  The QCD axion should be in the following
mass window
\cite{Preskill:1982cy,Abbott:1982af,Dine:1982ah,Chang:2018rso,Mayle:1987as,Raffelt:1987yt}:
\begin{align}
10^{-6}\EV\lesssim  m_a\lesssim 0.1\EV,
\end{align}
where the lower limit is from the mass density of the coherent
oscillation of the axion assuming that the initial misalignment angle
$\theta_i$ is of $O(1)$,\footnote
{Here we have assumed $H_{\rm inf}\gg 1\GEV$. If $H_{\rm inf}\ll
  1\GEV$ and inflation lasts long enough, $|\theta_i|\ll1$ follows
  from the equilibrium distribution during inflation, and this bound
  disappears~\cite{Graham:2018jyp, Guth:2018hsa}.}
while the upper limit is from the duration of neutrino burst in
SN1987A. By using misalignment mechanism to obtain the axion DM with
$m_a\gg 10^{-6}\EV$, $\theta_i$ should be fine-tuned so that the axion
initially stays (almost) at the top of the potential to enhance the
abundance due to an anharmonic
effect~\cite{Bae:2008ue,Visinelli:2009zm}. In such a case, the power
spectrum of the axion isocurvature perturbation is significantly
enhanced~\cite{Lyth:1991ub,Kobayashi:2013nva} and the isocurvature
problem, as well as the domain-wall problem~\cite{Takahashi:2019pqf},
may be serious. In particular, for $m_a\gtrsim 10^{-3}\EV,$ to evade
the isocurvature bound, the energy density during inflation is smaller
than $\MEV^4$, which is not consistent with the big-bang
nucleosynthesis (BBN).

The DASER production mechanism gives a new possibility to realize
axion DM with $m_a\gg 10^{-6}\EV$.  To realize the value of $\bar{f}$ in
Eq.\ \eqref{fitfbar}, the model parameters are related as
\begin{align}
  2\bar{f}\sim 50 \times
  \(\frac{1.7\times
    {10^{14}}\GEV}{\Lambda_a}\)^2 \left( \frac{r}{10^5} \right)
  \left( \frac{T_R}{10^5\ {\rm GeV}} \right),
  \label{pred1}
\end{align}
where, in this calculation, an (almost) instantaneous reheating is
assumed: $\rho_\f(t_i)\sim\frac{g_\star \pi^2 }{{30}}T_R^4$.  If
$t_R\ll t_i$, the required $\L_a$ becomes larger.  Therefore even if
the inflaton coupling is weak, $B$ can be large enough to have a
dominant component of the axion DM from the DASER production
mechanism.

 As we discuss in Section.\,\ref{sec:formalisms}, our
analysis is valid when the timescale of the dissipation is longer than
$\sim (qm_\phi)^{-1}\sim \L_a /(\bar{\f}m_\f)$ (see Eq.\,\eqref{phibardot} with $H\sim q^2 m_\f$ for DASER production).  In the case of the
instantaneous reheating, $t_R\sim t_i\sim 1/\G_{\rm dis}$, with
$\G_{\rm dis}\gtrsim H(t_R)$, where $\G_{\rm dis}$ is the dissipation
rate for the reheating.  Thus, we require $\G_{\rm dis}\lesssim
\bar{\f}m_\f/\Lambda_a$.

The DASER axion from the inflaton decay does not have an isocurvature
problem. This is because the axion density fluctuation follows that of
the inflaton.  In the model of Section \ref{sec:M2}, the axion DM in
the range of $10^{-4} \EV \lesssim m_a\lesssim 10^{-1}\EV$ can be
generated.  Such axion can give signal in DM
haloscopes~\cite{TheMADMAXWorkingGroup:2016hpc, Brun:2019lyf,
  McAllister:2017lkb, Marsh:2018dlj, Chigusa:2020gfs}, IAXO
experiment~\cite{Irastorza:2011gs, Armengaud:2014gea,
  Armengaud:2019uso}, and ARIADNE experiment~\cite{Arvanitaki:2014dfa,
  Geraci:2017bmq}.

The DASER axion may be thermalized due to the scattering with the
ambient plasma of the SM particles.  Assuming that the axion only
couples to the gluon (except for the inflaton) below the PQ scale, we
obtain the dissipation rate which is suppressed due to the nature of
derivative couplings (by neglecting a QCD sphaleron
contribution~\cite{McLerran:1990de}): $\Gamma_{\rm diss}^{\rm
  axion}\sim \frac{\a_s^2 T^3 }{32\pi^2 f^2_a} \frac{P_a^2}{g_s^4
  T^2}$, where $P_a$ is the typical momentum of the axion at SM
temperature, $g_s$ is the QCD coupling constant, and $f_a$ is the
axion decay constant \cite{Moroi:2014mqa}.  One finds that, for $T/P_a
\sim 2r\gg 1$, the rate is much smaller than the expansion rate $H$ if
\begin{align}
  \frac{T}{f_a}\lesssim 0.1 \times
  \(\frac{f_a}{10^9\GEV}\) \(\frac{r}{100}\)^{2}.
\end{align}
In the parameter region with $T_R \ll f_a$ and $r\gg 100$, the DASER
axion is not thermalized.\footnote
{If $T_R \gtrsim f_a$, we should consider the effects of heavy
  particles like the PQ scalars and PQ fermions.  We do not consider
  such a case.}

The DASER mechanism is generic for bosonic particles which is weakly
coupled to inflaton and SM particles; the candidates include hidden
photon and axion-like particle DM.  For instance, one can also use the
DASER mechanism to produce the hidden photon or axion DM to explain
the XENON1T excess~\cite{Aprile:2020tmw, Takahashi:2020bpq, Alonso-Alvarez:2020cdv, Athron:2020maw, Bloch:2020uzh, Nakayama:2020ikz, An:2020bxd, Li:2020naa}.  Conversely, such a DASER
production process may overproduce stable (or long-lived) bosons which
may result in cosmological problems with dark matters or dark
radiations.

\subsection{Dirac sea DM of right handed neutrino}
\label{sec:fermi}

RHNs are well-motivated particle to give masses to active neutrinos
via the seesaw mechanism~\cite{Minkowski:1977sc, Yanagida:1979as, Glashow:1979nm, GellMann:1980vs, Mohapatra:1979ia}.  Denoting the
left- and right-handed neutrinos as $\nu$ and $N$, respectively, the
Lagrangian in the seesaw scenario contains the following terms:
\begin{align}
  {\cal L} \supset - \frac{M_N}{2} \bar{N}^c N
  - y_\nu v \bar{N} \nu + \mbox{h.c.},
\end{align}
where $v$ is the expectation value of the Higgs field.  Here, for
simplicity, we assume that only a single flavor of the RHN is
important for our discussion.  Then, $\nu$ can be regarded as a
superposition of electron neutrino $\nu_e$, muon neutrino $\n_\m$, and
tau neutrino $\n_\t$:
\begin{align}
  \nu=\e_e \nu_e+\e_\m \nu_\m+ \e_\t \n_\t,
\end{align}
where $\e_i$'s, which are assumed to be real, denote coefficients
satisfying $\sum_{i={e,\m,\t}} \e_i^2=1$.  In the following, we focus
on the case of $M_N\gg y_\nu v$.  Then, by integrating out $N$, we
obtain an active neutrino mass term:
\begin{align}
  {\cal L}_{\rm eff} \supset \frac{(y_\nu v)^2}{2 M_N} \bar{\nu^c}\nu
  \equiv
  \frac{m_\nu}{2} \bar{\nu^c}\nu.
\end{align}
In this case, the mixing angle between the left- and right-handed
neutrinos is given by
\begin{align}
  \theta \equiv \frac{y_\nu v}{M_N}.
\end{align}
From the data of neutrino-oscillation experiments, we can estimate
\cite{Zyla:2020zbs}
\begin{align}
  m_\nu = \theta^2 M_N \sim (1 - 100) \ {\rm meV}.
  \label{m_nu}
\end{align}

\begin{figure}[!t]
\begin{center}  
   \includegraphics[width=145mm]{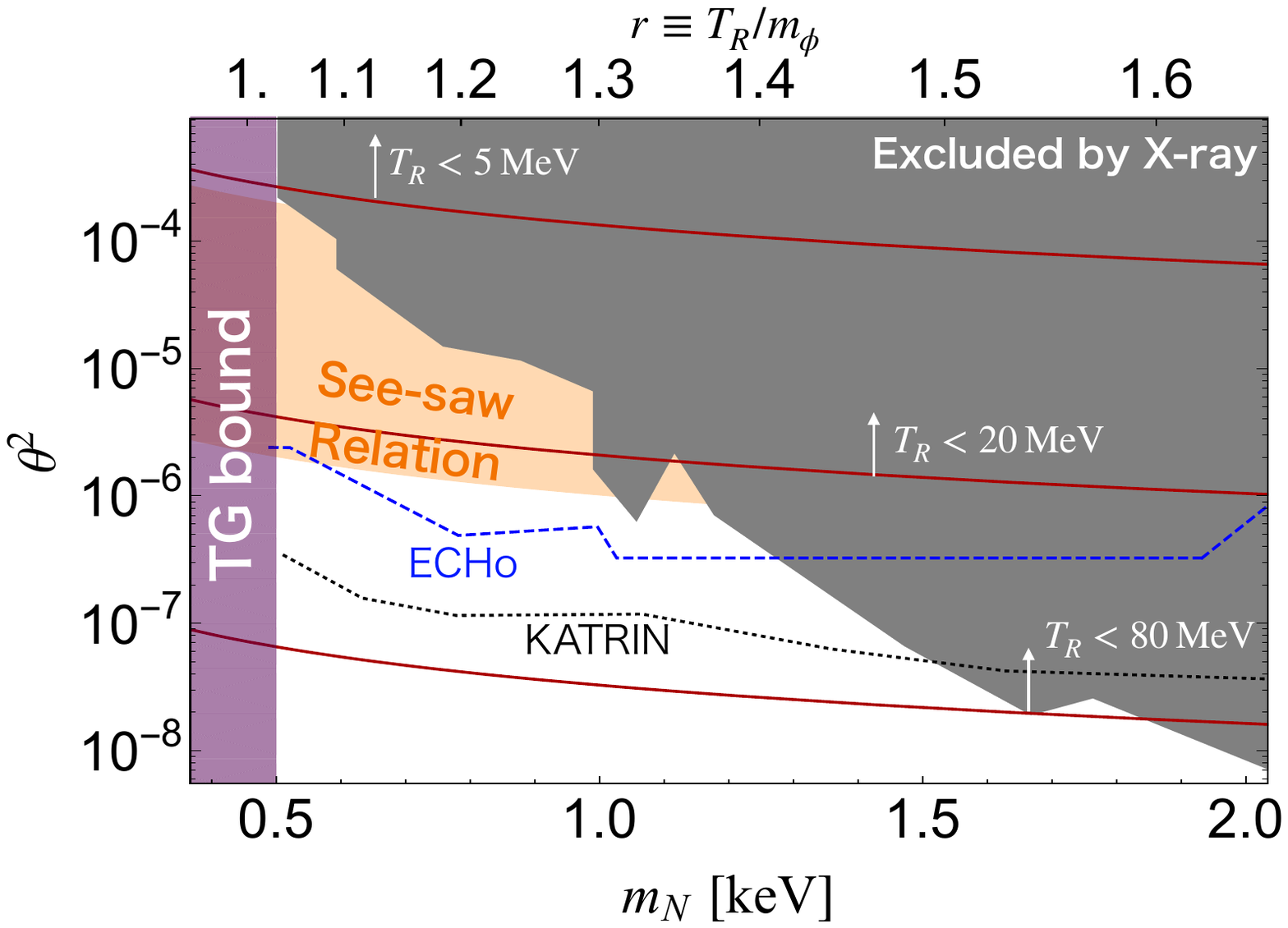}
      \end{center}
\caption{Parameter region for a Dirac sea DM of RHN in
  $m_N\text{-}\theta^2$ plane.  On the upper axis, we also show the
  predicted $r=T_R/m_\f$ for the Dirac sea DM with $g=2,
  g_{\star}=g_{\star, s}=11.$ Above the red solid lines, the reheating
  temperature should be smaller than $T_R=5\MEV, 20\MEV, 80\MEV$.}
\label{fig:2}
\end{figure}

The RHN (which may be also called as ``sterile neutrino'') $N$ is a
well-studied DM candidate~\cite{Asaka:2005an, Asaka:2005pn} (see also \cite{Adhikari:2016bei}).  If $N$ carries the momenta comparable to
the temperature of the active neutrinos, the Lyman-$\alpha$ constraint
forces the right-handed neutrino mass $M_N$ to be heavier than $2-
5\ {\rm keV}$ as a warm DM~\cite{Viel:2005qj, Irsic:2017ixq} (than
$5-20\KEV$ as a FIMP DM \cite{Kamada:2019kpe}).  For the RHN mass of
our interest, the RHN decays as $N\rightarrow\gamma \nu$ with
non-vanishing mixing angle $\theta$, and hence the mixing angle
$\theta$ is required to be highly suppressed to evade the bound from
$X$-ray observations~\cite{Adhikari:2016bei}.  It imposes a stringent
constraint on the scenario of the RHN DM \cite{Adhikari:2016bei}.
However, in our scenario, the RHN DM with sub keV mass may be produced
as Dirac sea and can be the dominant DM component without conflicting
the $X$-ray constraint as we see below.

Dirac sea of the RHN can be produced from the inflaton decay.  The
inflaton can have a coupling of the following form:
\begin{align}
  {\cal L} \supset y_\f \f \bar{N^c}N , 
\end{align}
where $y_\f$ is a dimensionless Yukawa coupling constant.  The decay
rate is
\begin{align}
\G_{\f\to NN }^{(0)} =  \frac{y_\f^2}{4\pi} m_\f.
\end{align}
If $y_\f$ is large enough, soon the decay process is Pauli-blocked.
Then the decay products form the Dirac sea.  Consequently the DM
abundance can be explained with $r\sim 1-10$ for $m_N=0.1-100\KEV$
(see the blue dashed line of Fig.\ \ref{fig:DM}).

In Fig.\,\ref{fig:2}, we show various constraints on the Dirac sea DM
of the RHN on $M_N$ vs.\ $\theta^2$ plane. In the orange band, the
see-saw relation given in Eq.\ \eqref{m_nu} is possible.  The $X$-ray
and TG bounds are taken from \cite{Adhikari:2016bei}.  The
sensitivities of the ECHo and KATRIN experiment are applicable to the
case that the RHN dominantly mixes with the electron neutrino (i.e,
$\e_e\sim 1$); if $0.5 \KEV \lesssim M_N\lesssim 2\KEV$, the existence
of the RHN can affect the shape of the deexcitation spectrum for
$^{163}$Ho by capturing an electron in ECHo~\cite{Filianin:2014gaa}
(see also HOLMES~\cite{Alpert:2014lfa} and NuMECS~\cite{Engle:2013qka}
experiments) and the shape of the tritium $\beta$ decay spectrum in
KATRIN~\cite{Angrik:2005ep, Mertens:2014nha,Mertens:2014osa} (see also
Troitsk \cite{Kraus:2004zw}, Project 8~\cite{Asner:2014cwa}, and
Ptolemy~\cite{Betts:2013uya} experiments).  When $\e_e<1$, the
sensitivity reaches on $\theta^2$ becomes worse by a factor of $\sim
1/\e_e^2$.  Thus, the Dirac sea DM of the RHN with $M_N\sim
0.5\text{-}1\KEV$ can be a viable DM candidate, avoiding both the TG
and $X$-ray bounds.\footnote
{These bounds could be further alleviated if we have more than one RHN
  to be dominant DM.}

In order for the Dirac sea RHN DM, the RHN should not be thermalized.
The thermalization rate is estimated as $\G_{\rm th}^N=\theta^2
\G_{\rm th}^\nu,$ where $\G_{\rm th}^\nu\sim G_F^2 T^5$ is that of an
active neutrino with $G_F$ being the Fermi constant.\footnote
{Notice that, when $r\gg 1$, there would be a further suppression of
  $1/r$ for the scattering rate since the center-of-mass energy would
  be $\sim \sqrt{1/r} T$.  For the case of our interest, however,
  $r\sim1$ and we neglect such a correction.}
It is known that the active neutrino decouples from the thermal bath
at the temperature below $\sim 1\MEV$.  Using the fact that the
upper bound can be obtained by solving $H\sim\G_{\rm th}^N$, the upper
bound from the non-thermalization is estimated as $T_R \lesssim 1\,
{\rm MeV}\times\theta^{-2/3}$.  A more stringent bound comes from the
thermal production of the RHN, whose abundance can be estimated as
\cite{Gelmini:2004ah}
\begin{align}
  \Omega_N^{\rm th} h^2 \sim 0.11 \( \frac{\sin^2{2\theta}}{10^{-3}}\) 
  \(\frac{m_N}{1\KEV}\) \(\frac{T_R}{5\MEV}\)^3.
\end{align}
Here again we assumed $\epsilon_e=1.$ As previously noted, the thermal
component of the RHN should not be dominant for $m_N\lesssim 10\KEV$
to evade the Lyman-$\a$ bound.  In Fig.\,\ref{fig:2}, we show the
contours for $\Omega_N^{\rm th} h^2 = 0.12/2$.  Above the red solid
lines to satisfy the condition, the reheating temperature should
satisfy $T_R=5, 20,80\MEV$ from the top to the bottom.  Thus, even if
the reheating temperature is much higher than $\sim 1\ {\rm MeV}$ in
order not to affect the BBN, we can avoid the overproduction of the
RHN from the thermal production.

Interestingly, the Dirac sea RHN DM satisfying the see-saw relation
not only can be searched for but also provide implication of the early
Universe,
\begin{align}
\O( 1)\MEV \lesssim m_\f \sim T_R \lesssim \O(10)\MEV.
\end{align}
In particular with $\e_e^2=O(0.1-1)$ the scenario can be tested
in KATRIN and ECHo.

An interesting possibility of our scenario may be inflaton
hunts~\cite{Bezrukov:2009yw, Takahashi:2019qmh, Okada:2019opp}.  In
order for the present scenario, the inflaton should be as light as
$1-100\MEV$ and couple to SM particles for reheating.  Therefore, it
could be produced by current or future experiments.  For instance, if
we consider that the reheating is due to the interaction via inflaton
mixing with the Higgs field, the inflaton with $m_\f\lesssim 100\MEV$
dominantly decays as $\f \to e^+ e^-$ at the rate of $\G^{0}_{\f\to
  e^+e^-}\sim \theta_H^2 \frac{m_e^2 }{16\pi v^2} m_\f$, with
$\theta_H$ being the inflaton-Higgs mixing angle.
Reheating temperature due to this decay process is estimated as\footnote
{Since $r\sim 1$, we can neglect the thermal back reaction discussed
  in Section\,\ref{sec:M2}.} 
\begin{align}
  T_R\sim5\MEV \sqrt{\frac{m_\f}{5\MEV}} \(\frac{\theta_H}{2 \times 10^{-4}}\)\( \frac{10}{g_{\star}}\)^{1/4}.
\end{align} 
from $3H(T_R)\sim \Gamma_{\f\to e^+e^-}.$
This satisfies the constraint $\theta_H<3\times 10^{-4}$ from the $K$ meson
decays~\cite{Artamonov:2009sz, Winkler:2018qyg} and may be searched
for in the future experiments~\cite{Batell:2019nwo,CortinaGil:2020vlo}.

\section{Reheating from Evaporating Inflaton}
\label{sec:M2}
\setcounter{equation}{0}

In this section, we present an example of the model which gives rise
to $r\gg1$.  During $t_i \lesssim t\lesssim t_R$, the reheating takes
place with transferring the energy density of the inflaton to that of
the SM particles.  Here, in order for the reheating process, we
consider the interaction between inflaton and SM particle with the
following
higher dimensional operator:
\begin{align}
{\cal L}\supset \frac{\f }{\L_G} G_{\mu\nu}^{(a)} G^{(a)\mu\nu},
\end{align}
where $G_{\mu\nu}^{(a)}$ is the field strength of gluon and
$\Lambda_G$ is the cut-off scale.  This operator could be generated if
heavy extra quarks couple to $\f$.  (If the extra quarks have quantum
numbers of $SU(2)_L$ and $U(1)_Y$ gauge interactions, $\phi$ may also
couple to those gauge fields.  For simplicity, however, we adopt the
interaction given above for our argument.)  Then, the perturbative
decay rate of $\phi$ in the vacuum is given by 
\begin{align}
  \Gamma_{\phi\rightarrow GG}^{(0)} \simeq \frac{2m_\phi^3}{\pi \Lambda_G^2},
\end{align}
where we have implicitly assumed that $m_\phi$ is heavier than
the QCD scale,
\begin{align}
\label{ineq1}
m_\f\gtrsim 0.1\GEV,
\end{align}
so that the non-perturbative effects from QCD dynamics is unimportant.

If the rate is large enough, the decay with a Bose-enhancement
produces numerous gluons. Unlike the DM, the gluons have strong
interaction and ambient plasma of the SM particle with temperature $T$
is soon formed.  The enhancement is then disturbed by the thermal
effect; it could be interpreted as the kinematical block due to the
thermal mass of the gluon, which is of order $g_s T$.  Assuming that
the thermalization processes are so fast that the cosmic expansion is
unimportant for its study, the energy density of the radiation,
denoted as $\rho_r$, evolves as $\dot{\r_r}\sim -\G_{\f \to
  GG}^{(0)}\rho_\f$.  Thus, the timescale relevant for the 
effects of the thermal blocking is estimated as
\begin{align}
  t^{(\rm block)} \sim
  \frac{\rho_r|_{T\sim m_\f/g_s}}{\G_{\f \to GG}^{(0)}\r_\f}.
\end{align}
This timescale is shorter than the cosmic time if
\begin{align}
\label{block}
\L_G \lesssim 1.3\times 10^{14}\GEV \(\frac{100\GEV}{m_\f}\)^{1/2} \(\frac{\rho_\f^{1/4}}{100\GEV}\).
\end{align}

Even if the kinematical block occurs, the dissipation of $\phi$ may
proceed due to the multiple scattering of $\f$ in the thermal plasma
\cite{Laine:2010cq, Moroi:2014mqa}.  From the dimensional analysis, the
dissipation rate is estimated as
\begin{align}
  \G_{\text{dis},G} \sim C \frac{T^3}{\L_G^2}.
\end{align}
where $C$ is a numerical constant.  In \cite{Laine:2010cq}, $C$ is
estimated from the imaginary part of the two point function of $\phi$:
\begin{align}
  C \sim \frac{(12\pi\a_s)^2}{\log{(1/\a_s)}},
\end{align}
with $\a_s=\frac{g_s^2}{4\pi}$.  One can find that the dissipation
effect becomes important at higher temperature.  This implies that
the reheating via the dissipation effect may become important soon
after the inflation (preheating).  As discussed in
\cite{Laine:2010cq}, the dissipation rate can hardly win the expansion
rate of the universe if $ \Lambda_G\sim M_{\rm pl}$.  In the
following, we consider the case that $\Lambda_G\ll M_{\rm pl}$ and
the dissipation effect plays an important role in the reheating.

In discussing scenarios with dissipation, it should be noted that the
dissipation rate is of the same order of the thermal production rate
of the inflaton (as far as the cosmic temperature is higher than
$m_\phi$); both of them are of the order of $\sim T^3/\L_G^2$, by
using the fact that the QCD coupling constant is sizable.  Thus, if
the dissipation rate becomes larger than the expansion rate of the
universe, $\phi$ is likely to be thermalized.  Such a thermally
produced inflaton may affect the BBN processes in particular if the
decay of the inflaton happens at the cosmic temperature of $\sim
1\ {\rm MeV}$ or lower.  (See \cite{Kawasaki:1999na, Kawasaki:2000en,
  Hannestad:2004px, Ichikawa:2006vm, DeBernardis:2008zz,
  deSalas:2015glj, Hufnagel:2018bjp, Hasegawa:2019jsa, Kawasaki:2020qxm, Depta:2020zbh} for the BBN
bounds on long-lived particles.)  In order that the thermally produced
inflaton does not affect the standard BBN predictions, we require that
the inflaton should disappear from the thermal bath well before the
BBN epoch (i.e., conservatively, at $T\gtrsim 10\ {\rm MeV}$).  Such a
requirement can be satisfied if
\begin{align}
  & T_{\phi{\rm \mathhyphen decay}}\gtrsim 10\MEV,
  \label{ineq2}
  \\
  &
  m_\phi \gtrsim  10\MEV.
\end{align}
Here, $T_{\phi{\rm \mathhyphen decay}}$ is the cosmic temperature just
after the decay of the inflaton (assuming that the inflaton decays
after becoming non-relativistic); $T_{\phi{\rm \mathhyphen decay}}$ is
estimated by solving $3H(T_{\phi{\rm \mathhyphen decay}})\sim \G_{\f
  \to GG}$, and is obtained as
\begin{align}
  T_{\phi{\rm \mathhyphen decay}} \sim
  \left( \frac{g_\star \pi^2}{90} \right)^{-1/4}
  \sqrt{\Gamma_{\f \to GG} M_{\rm pl}/3}.
\end{align}
Note that the thermally produced inflaton may dominate
the Universe. In such a case, $T_{\phi{\rm \mathhyphen decay}}$ is
regarded as the reheating temperature due to the decay of such
inflaton.

We can check numerically whether the reheating is successful.
We solve the following set of Boltzmann equations:
\begin{align}
  \label{Bolt1}
  \dot{\rho}_\f+3H \rho_\f & = -
  ( \Gamma_{\phi\rightarrow GG}^{(T)} + \G_{\rm dis,G} ) \rho_\f,
  \\
  \label{Bolt2}  \dot{\rho}_r+4H \rho_r & =
  ( \Gamma_{\phi\rightarrow GG}^{(T)} + \G_{\rm dis,G} ) \rho_\f,
\end{align}
where $\Gamma_{\phi\rightarrow GG}^{(T)}$ is the decay rate of $\phi$
in thermal bath.  The detailed study of the effects of the thermal
blocking is beyond the scope of this paper.  We expect that the
perturbative decay proceeds at least until the cosmic temperature
becomes as large as $\sim m_\f/g_s$.  Then, for $T\gtrsim m_\f/g_s$,
the perturbative decay of $\phi$ is suppressed because of the thermal
blocking.  As an approximated procedure, in our numerical analysis, we
solve the above set of Boltzmann equations with the initial condition
of $\rho_r=C g_\star \pi^2 T^4/30|_{T=m_\f/g_s}$ (with $C$ being a
constant of $O(0.1-1)$) and $\rho_\f= g_\star \pi^2 T_R^4/30$, with
setting $\Gamma_{\phi\rightarrow GG}^{(T)}=0$ (assuming that the
perturbative decay is maximally blocked for $\rho_r>C g_\star\pi^2
T^4/30|_{T=m_\f/g_s}$).  In our analysis, we concentrate on the case
that the narrow resonance condition for the gluon production process
holds, leaving the study of the case of the broad resonance for a
future work.  Such a condition is given by $q'\equiv
\frac{\bar{\phi}}{\Lambda_G }\lesssim 1$.  Requiring $q'\lesssim 1$ at
the time of the reheating, we obtain
\begin{align}
  \Lambda_G \gtrsim 10^{10}\GEV \times
  \(\frac{ r}{10^5}\)^2 \(\frac{m_\f}{1\GEV}\).
  \label{ineq3}
\end{align}
In this regime, the expansion rate during the inflation $H_{\rm inf}$
may satisfy $\G_{\text{dis},G}\gtrsim H_{\rm inf}$, with which a
successful instantaneous reheating occurs.

\begin{figure}[!t]
  \begin{center}  
    \includegraphics[width=145mm]{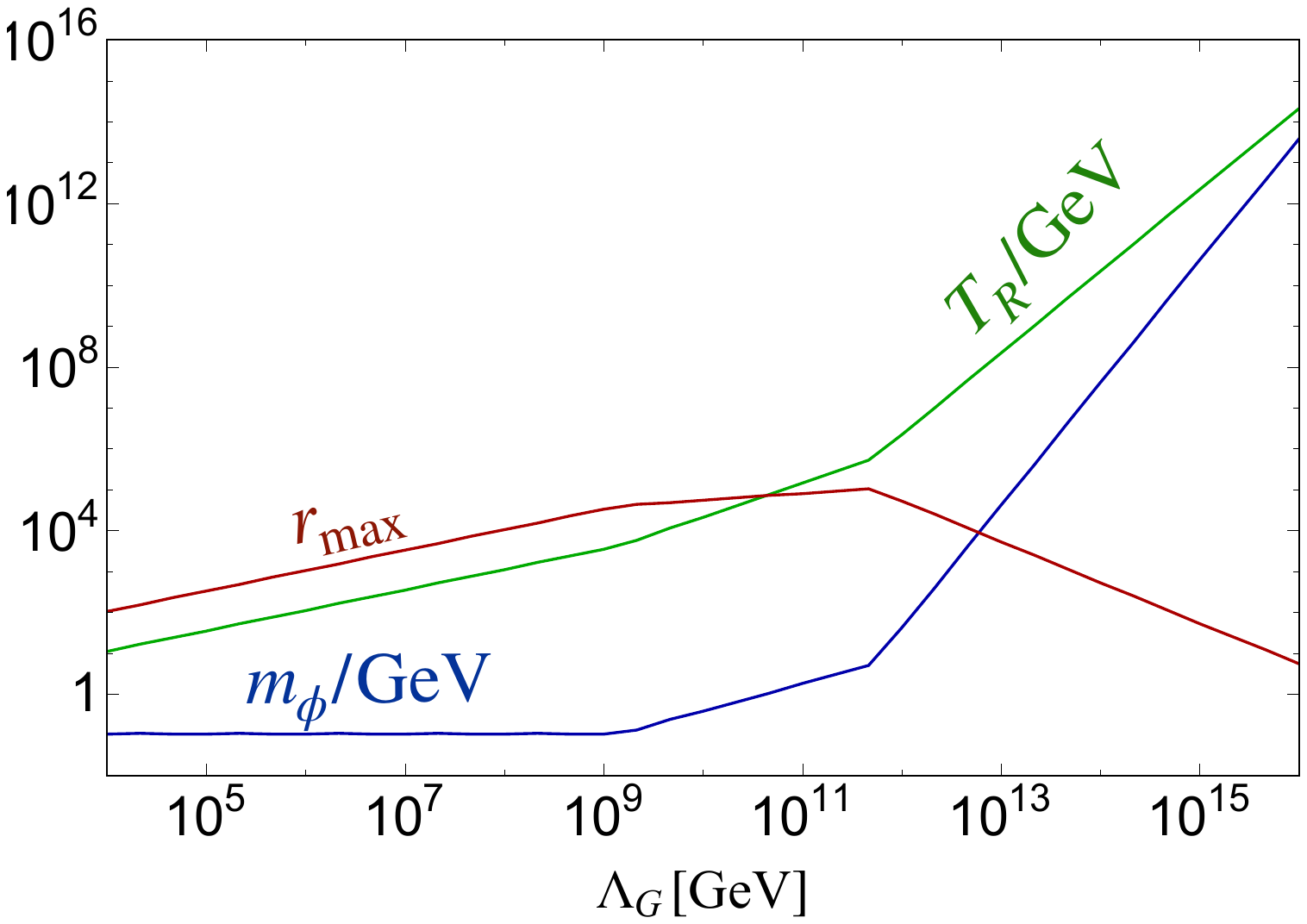}
  \end{center}
  \caption{ The maximal $r$ by varying $\L_{G}$ [red line]. The line
    satisfies the inequalities of \eqref{osccon}, \eqref{block},
    \eqref{ineq1}, \eqref{ineq2}, and \eqref{ineq3}. The corresponding
    reheating temperature and inflaton mass are also shown in green
    and blue solid lines, respectively.}
  \label{fig:reh2}
\end{figure}

We numerically solve the set of Boltzmann equations \eqref{Bolt1} and
\eqref{Bolt2} and find that the energy density of $\phi$ can be
efficiently converted to that of radiation for some region of the
parameter space.  In Fig.\,\ref{fig:reh2}, we show the maximal value
of $r$, $r_{\rm max}$, as well as the corresponding $m_\f$ and $T_R$
by varying $\L_G$.  The lines satisfy the inequalities \eqref{osccon},
\eqref{ineq1}, \eqref{block}, \eqref{ineq2}, and \eqref{ineq3}.  In
the present model, we find that $r$ can be as large as $10^5-10^6$.
We see that the slope suddenly changes at around $\L_G=10^9\GEV$ and
$10^{11-12}\GEV$.  For $\L_G<10^9\GEV$
($10^9\GEV<\L_G<10^{11-12}\GEV$, $\L_G>10^{11-12}\GEV$), $r_{\rm max}$
is determined by the constraints \eqref{ineq1} and \eqref{ineq2}
(\eqref{ineq2} and \eqref{ineq3}, \eqref{osccon} and \eqref{ineq3}).
The constraint \eqref{block} is less stringent than the others in the
whole range.  The maximal value of $r$ can be obtained when $\L_G\sim
10^{10-11}\GEV,$ $T_\f\sim 10\MEV$ and $\bar{\f}/\L_G\sim 1$.  We have
checked that the result is insensitive to the choice of $C$ as far as
$C\sim 0.1-1$.  Thus, the reheating temperature higher than the
inflaton mass is possible in the present scenario.

As we mentioned, the thermally produced inflaton may dominate the
Universe subsequently and then decay.  The DM velocity is further
suppressed by the extra dilution due to the decay of the thermally
produced inflaton.  The dilution factor $\D$ is estimated as
$\D\sim T_{\rm \f\text{-}decay}/m_\f$ and, numerically,
\begin{align}
  \D^{1/3}\sim 0.2 \times
  \(\frac{m_\f}{1\GEV}\)^{1/6}
  \(\frac{10}{g_\star}\)^{1/12}\(\frac{10^{11}\GEV}{\L_G}\)^{1/3}.
\end{align}
With the effect of the extra dilution due to the decay of thermally
produced inflaton, the Lyman-$\alpha$ bound in Fig.\,\ref{fig:DM} can
be alleviated and $m_\chi\sim 10^{-4}\EV$ may become possible without
conflicting the Lyman-$\a$ constraint.  We also emphasize that,
although an extra entropy production can make the DM colder, we still
need $r\gtrsim 1$ to make the DM number density large enough.

Again, from Eq.\,\eqref{phibardot} and $H \sim q^2 m_\f$, we need $ q
m_\phi \gtrsim \Gamma_{{\rm dis}, G}$ for the consistency of the DM
production, which implies
\begin{align}
q \gtrsim 10^{-12}\(\frac{m_\f}{1\GEV}\)^2 \(\frac{10^{11}\GEV}{\Lambda_G}\)^2 \(\frac{r}{10^3}\)^3.
\end{align}
For the axion DM, we have $q\sim \bar{\f}/\Lambda_a \sim 10^{-3} q'$ for $\Lambda_a/\Lambda_G\sim 10^{3}.$ 
Thus, this can be easily satisfied. 

\section{Conclusions and discussion}
\label{sec:conclusions}
\setcounter{equation}{0}

We have studied the DM production due to the decay of the inflaton (or
more generically, an oscillating scalar field).  If the inflaton
coupling to the SM sector has a sizable strength, the reheating
temperature due to the inflaton decay can be comparable to or higher
than the inflaton mass.  If $T_R\gtrsim m_\phi$, then the DM produced
by the inflaton has a momentum smaller than those of particles in the
thermal bath (consisting of the SM particles), assuming that the
interaction of the DM with the SM particles or itself is negligibly
weak. It can help the DM produced by the decay to be cold enough to be
consistent with the Lyman-$\alpha$ bound on warm DMs even if the DM
mass is smaller than $O(1)\ {\rm keV}$.

We have also shown that, if the DM is bosonic, the production of the
DM from the inflaton decay can be enhanced due to the effect of the
stimulated emission, like the LASER.  The mechanism, called DASER
(i.e, the DM amplification by stimulated emotion of radiation), can
significantly enhance the DM abundance and can make light bosonic DM
scenarios viable.  In addition, if the DM is fermionic, the DM
produced by the decay may form Dirac sea.  The DASER and Dirac sea DMs
produced by the inflaton decay may be searched for by future
observations of the 21cm lines.  The DMs produced by the above
mentioned mechanism have null isocurvature perturbations.

Finally, we comment that the DMs produced in the present scenarios may
have very special momentum distribution, and the information about the
production mechanism discussed here may be embedded in the momentum
distribution of the DM.  For example, the momentum of the bosonic DM
produced by the DASER mechanism would have sharp peak at the IR mode
at the momentum $\sim m_\f/a[t_i].$ From a quantum field theory
approach, the peak width can be approximated as $\sim \sqrt{m_\f H
  (t_i)}/a(t_i)$ (see \cite{Moroi:2020bkq}).  By combining the two, we can
get $m_\f/H(t_i).$ In addition, momentum distribution of the modes
produced during the preheating depends on the thermal history during
the preheating.  One may probe the reheating phase (see also
Ref.\,\cite{Jaeckel:2020oet}) or preheating phase if the information
about the momentum distribution of the DM becomes available.  For this
purpose, further study of the structure formation with non-standard
momentum distribution of the DM is needed.

\section*{Acknowledgement}
This work is supported by JSPS KAKENHI grant Nos.\ 16H06490 (TM and
WY) and 18K03608 (TM).

\end{document}